\documentclass[12pt,preprint]{aastex}

\newcommand{\ratioo}{\rm {N}({\rm H}_2) / {\rm {I}}_{\rm CO}}

\shorttitle{Molecular Gas in the Outskirts of NGC~6946}
\shortauthors{Braine et al.}

\begin{document}


\title{The Detection of Molecular Gas in the Outskirts of
NGC~6946}


\author{Jonathan Braine\altaffilmark{1}, Annette M.~N. Ferguson\altaffilmark{2}, Frank Bertoldi\altaffilmark{3}, Christine D. Wilson\altaffilmark{4}}

\altaffiltext{1}{Laboratoire d'Astrophysique de Bordeaux, Universit\'e Bordeaux 1, Observatoire
   de Bordeaux, OASU, UMR 5804, CNRS/INSU, B.P. 89, Floirac, F-33270,
   France, braine@obs.u-bordeaux1.fr}
\altaffiltext{2}{Institute for Astronomy, University of
   Edinburgh, Royal Observatory, Blackford Hill, Edinburgh, EH9 3HJ,
   UK, Ferguson@roe.ac.uk}
\altaffiltext{3}{Argelander-Institut f\"{u}r Astronomie,
   Universit\"{a}t Bonn, Auf dem H\"{u}gel 71, Bonn, D-53121, Germany, bertoldi@astro.uni-bonn.de}
\altaffiltext{4}{Department of Physics \& Astronomy, McMaster
   University, Hamilton, Ontario, L8S 4M1 Canada, wilson@physics.mcmaster.ca}


\begin{abstract}
  We present the results of a search for molecular gas emission via
  the CO line in the far outer disk of the nearby spiral, NGC~6946.
  The positions targeted were chosen to lie on or near
  previously-identified outer disk HII regions. Molecular gas was
  clearly detected out to 1.3~R$_{25}$, with a further tentative
  detection at 1.4~R$_{25}$. The CO detections show excellent
  agreement with the HI velocities and imply beam-averaged column
  densities of $0.3-9\times 10^{20}$~cm$^{-2}$ and molecular gas
  masses of (2-70)$\times 10^{5}$~M$_{\sun}$ per 21$''$ beam (560pc).  
  We find evidence for an
  abrupt decrease in the molecular fraction at the edge of the optical
  disk, similar to that seen previously in the azimuthally-averaged
  areal star formation rate.  Our observations provide new constraints
  on the factors that determine the presence and detectability of
  molecular gas in the outskirts of galaxies, and suggest that neither
  the HI column, the metallicity or the local heating rate alone plays
  a dominant role.
\end{abstract}


\section{Introduction}

It has become increasingly apparent in recent years that star
formation is not just confined to the optically-bright parts of
galaxies.  Signposts of low-level star formation, such as HII regions
and UV-bright clusters, have been discovered in the far outer regions
of galactic disks, well beyond the classical R$_{25}$ radius
(e.g. \cite{ferg98a,thilker05}), and in the tidal HI filaments of some
interacting systems (e.g. \cite{ryanw04, Braine01}).  The existence of
star formation in these unusual environments, characterised by low
metallicity, low HI column density and low interstellar radiation
field, provides important constraints on star formation in
galaxies, as well as the necessary conditions for sustaining a
multi-phase interstellar medium (ISM).

Much theoretical effort has been devoted to understanding the 
formation of a cold molecular phase in the ISM, a prerequisite for 
star formation.  Thermal instabilities are generally believed
to be the primary mechanism for converting cool atomic gas into cold
molecular gas once a minimum pressure or local threshold surface
density is exceeded (e.g. \cite{elme94,schaye04, blitz06}). 
Additional mechanisms for forming unstable cold clouds include shocks
from spiral density waves and swing-amplifier and magneto-Jeans
instabiities (e.g. \citet{dobbs06, kim02}).  The threshold column
densities predicted for molecular cloud formation are on average a few
times smaller than those inferred for massive star formation in galaxies
(e.g. \cite{skill87,mk01}).

Direct detection of molecular clouds in extreme environments provides
a means to test ideas about cloud formation and survival.  Carbon
monoxide (CO) is the most abundant heteronuclear molecule (i.e. with
permitted rotational transitions) and is excited by collisions at
densities typical of molecular clouds, making CO the standard tracer
of molecular gas.  Concerns have previously been raised 
\citep[e.g. ][]{Pfeniger94} as to whether
the physical conditions in the outskirts of disks are sufficient to
excite CO to detectable temperatures and whether the metallicities in
these regions are sufficient to form enough CO for detection.  With
the discovery of CO emission out to 1.5~R$_{25}$ in the disk of
NGC~4414 \citep{braine04} and in high column density tidal HI filaments in the
M81 Group \citep{brou92,walter06}, these concerns have been proven
somewhat unfounded.

In this Letter, we present the results of a search for molecular gas
in the far outer disk of the nearby spiral NGC~6946.  While the study
of molecular gas in NGC~4414 focused on high column HI regions
\citep{braine04}, we have chosen to search for CO emission in the
vicinity of outer disk star-forming regions. NGC~6946 was previously
identified through H$\alpha$ observations to exhibit very extended
massive star formation, out to $\sim 2$R$_{25}$ \citep{ferg98a}, and thus
presents an excellent opportunity to attempt to measure and quantify
molecular gas at extreme radii.  We also report on the null detection
of molecular gas at two positions in the far outer disk of another
spiral with extended star formation, NGC~1058.

\section{Observations}

Our search was conducted using the IRAM
30m telescope at Pico Veleta, Spain during July 7-13 1999 in very good
summer weather.  The $^{12}$CO J$=1-0$ and J$=2-1$ transitions
at 115.271 and 230.538 GHz were observed simultaneously in two
polarizations (H and V) with the newly installed "AB" SIS receivers.
The filter bank back end was configured into four units of 256 x 1MHz
channels, one for each polarization and each frequency, yielding a 650
km s$^{-1}$ bandpass and a channel separation of $\sim 2.6$~km
s$^{-1}$ in the CO(1--0) transition, respectively 330 and 1.3~km
s$^{-1}$ in the CO(2--1) transition.  Typical system temperatures
(T$_a^*$) for the observations of NGC~6946 and NGC~1058 were 250~K at
115 GHz and 400~K at 230 GHz.  All spectra are presented on the
main beam temperature scale assuming forward and main beam
efficiencies of 0.9 and 0.73 at 115 GHz and 0.84 and 0.51 at 230GHz.

\begin{figure}
\begin{center}
\includegraphics[width=0.80\linewidth,angle=-90]{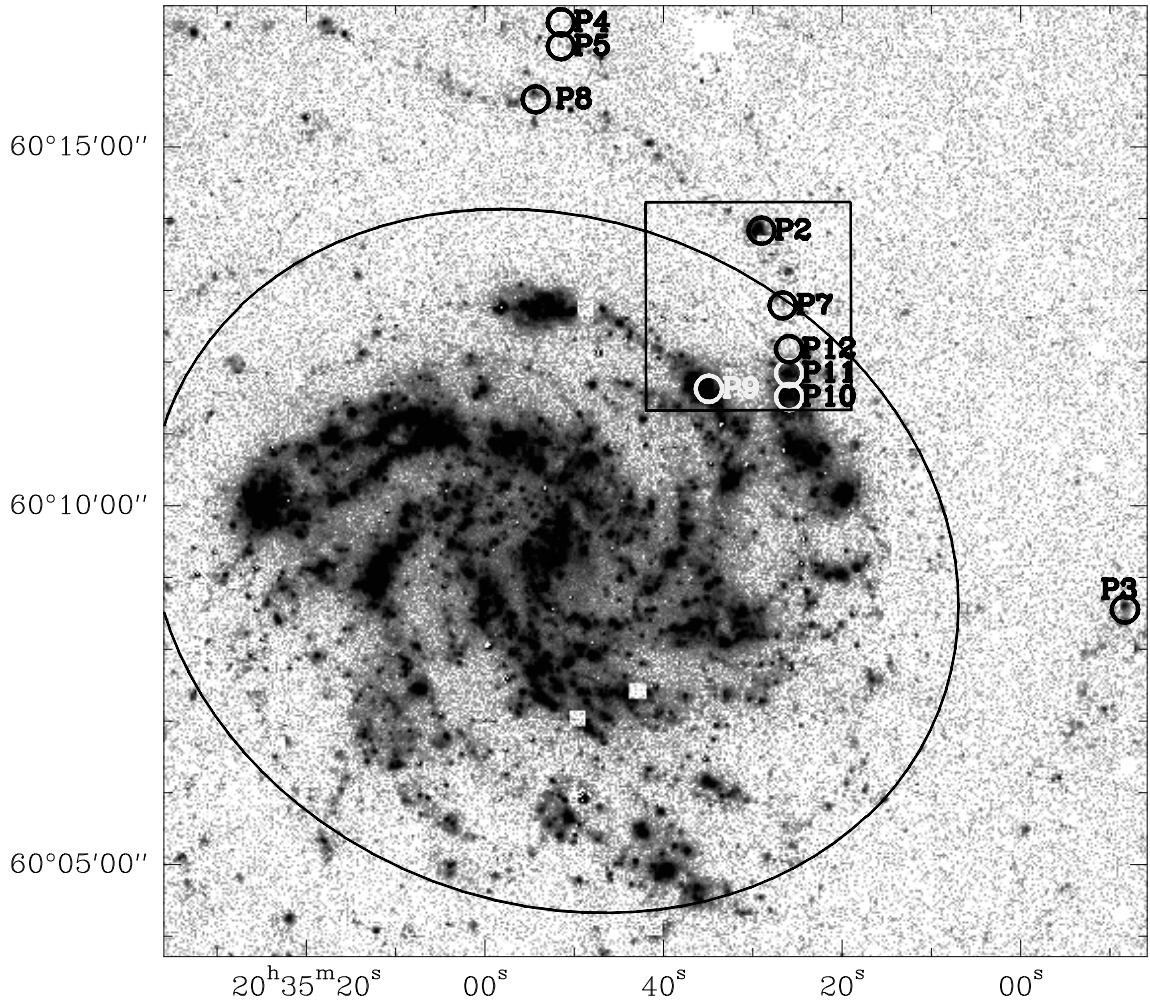}
\end{center}
\caption{An H$\alpha$ image of NGC~6946. The white and
black circles show the IRAM positions observed and the size is that of the CO(1--0)
beam. The black ellipse shows the R$_{25}$ contour. The box shows the region covered in Figure \ref{beams}.
\label{galaxy}}
\end{figure}

Single pointings were obtained on or near HII regions lying at radii
in the range 0.7--1.6~R$_{25}$ in NGC~6946 (see Figure~\ref{galaxy}),
and 1.0--1.9~R$_{25}$ in NGC~1058.  The IRAM 30m beam size is
21$\arcsec$ at 115 GHz and 11$\arcsec$ at 230 GHz, corresponding to
560 and 280 pc respectively at the distance of NGC~6946 (5.5~Mpc, see
\citet{regan06}) and 1~kpc and 0.5~kpc at the distance of NGC~1058
(10~Mpc, see \cite{ferg98a}).  As can be seen in Figure
{\ref{galaxy}}, the CO beam size was usually larger than the size of
the target HII regions.

After eliminating obviously bad spectra and interpolating over
occasional bad channels, spectra were summed for each transition and
position.  Despite observing in position-switching mode with a
reference position typically $\sim 3$ R$_{25}$ from the galaxy, only a
zero-order baseline subtraction was required.

\section{Results}

The results of our CO search are presented in Table 1.  In NGC~6946,
CO is clearly detected at 7 out of 10 positions, ranging from
0.7--1.3~R$_{25}$, with a further tentative detection at 1.4R$_{25}$
(P3).  In NGC~1058, no detections were made.

\begin{figure}
\begin{center}
\includegraphics[width=0.45\linewidth,angle=0]{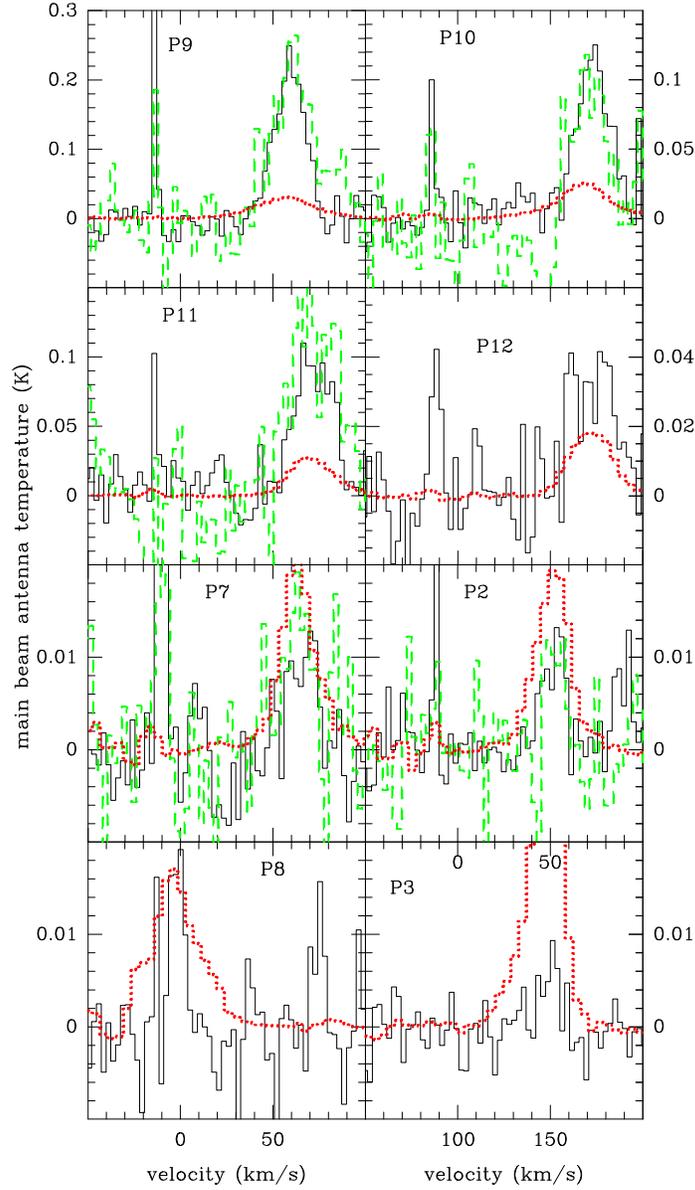}
\end{center}
\vspace*{2cm}
\caption{CO spectra of the detections in NGC~6946, along with P3 which
  is a tentative detection.  The vertical axis shows the main beam
  antenna temperature expressed in Kelvins and the horizontal axis the
  heliocentric velocity in km~s$^{-1}$. The CO(1--0) spectra are shown
  as solid black lines, the CO(2--1) as green dashed lines when
  detected, and the HI spectra as red dotted lines.  The HI scale is
  constant with respect to CO in order to show how
  relative HI and CO line strengths vary from position to position.
  Dividing the CO scale by 3 gives the HI flux in Jy~beam$^{-1}$.
  The narrow lines at $-15 \la v \la -8$ km~s$^{-1}$ are due to local
  emission from high-latitude Galactic clouds along the line of 
  sight (b $\sim 12^o$). Note the vertical scale
  varies between plots.  \label{spectra}}
\end{figure}

Figure~\ref{spectra} shows the CO(1-0) (and CO(2-1) when detected)
spectra of the eight positions in NGC~6946 with molecular gas
emission.  
CO(2--1) emission is detected at most positions with CO(1--0)
emission which suggests that the gas is not highly subthermally excited, as
it might be if it were extremely diffuse.
HI spectra extracted at the same positions and with the
same synthesized beam size are overlaid, taken from the 21cm data cube
of \cite{boomsma07}.  The HI and CO velocities show excellent
agreement, with the maximum offset being less than 3 km~s$^{-1}$.  The
CO line widths, as determined from gaussian fits, are in the range
8--30~km~s$^{-1}$ and are greater than those expected from individual
Giant Molecular Clouds (GMCs), even at the outermost positions.  This
implies that an ensemble of clouds is observed within the 560 pc
CO(1--0) beam in each case, similar to the findings of
\cite{braine04} for the outer disk of NGC~4414.

The spectra at the two outermost positions warrant further discussion.
The systemic velocity of NGC 6946 is very close to the velocity of the HI and CO in the Milky Way, thus making comparisons at some positions more difficult than they would be in higher velocity systems.
The detection at P8 (1.3~R$_{25}$) could be called into question since
the emission line is separated by only a single velocity channel from
the narrow Galactic line at this position.  
If the line emission were Galactic in origin
then we would expect it to have a significant spatial extent on the
sky and to be detected at positions P5 and P4. 
For a GMC lying 500~pc along the line of sight
(i.e. 100 pc above the Galactic plane), P5 and P4 correspond
to linear distances of 0.12 and 0.16~pc from P8.  No CO emission at the
velocity detected at P8 is seen at these positions.  
Fig. 2 shows that weak HI is detected at the local velocity
($\sim -15$ km~s$^{-1}$) by the WSRT in all positions but there is likely some 
spatial filtering out of the local HI signal.  At P8 the HI is much stronger at the velocity 
of the CO we attribute to NGC 6946, further strengthening our argument that the
P8 detection is real and that the line belongs to NGC 6946. 
We thus consider P8 to be a genuine detection of CO emission from NGC~6946.

The line emission at 150~km~s$^{-1}$ towards P3 (1.4~R$_{25}$) is
statistically significant ($\sim 5 \sigma$) but another "line" of
similar strength is seen at 270~km~s$^{-1}$.  The precise agreement
between the CO and HI line velocities at this position certainly
supports the reality of the CO feature but the presence of the other
peak casts some doubt.  As a result, we consider P3 as a probable but
still tentative detection.

\begin{figure}
\begin{center}
\includegraphics[width=0.6\linewidth,clip, trim = 30mm 60mm 0mm  0mm]{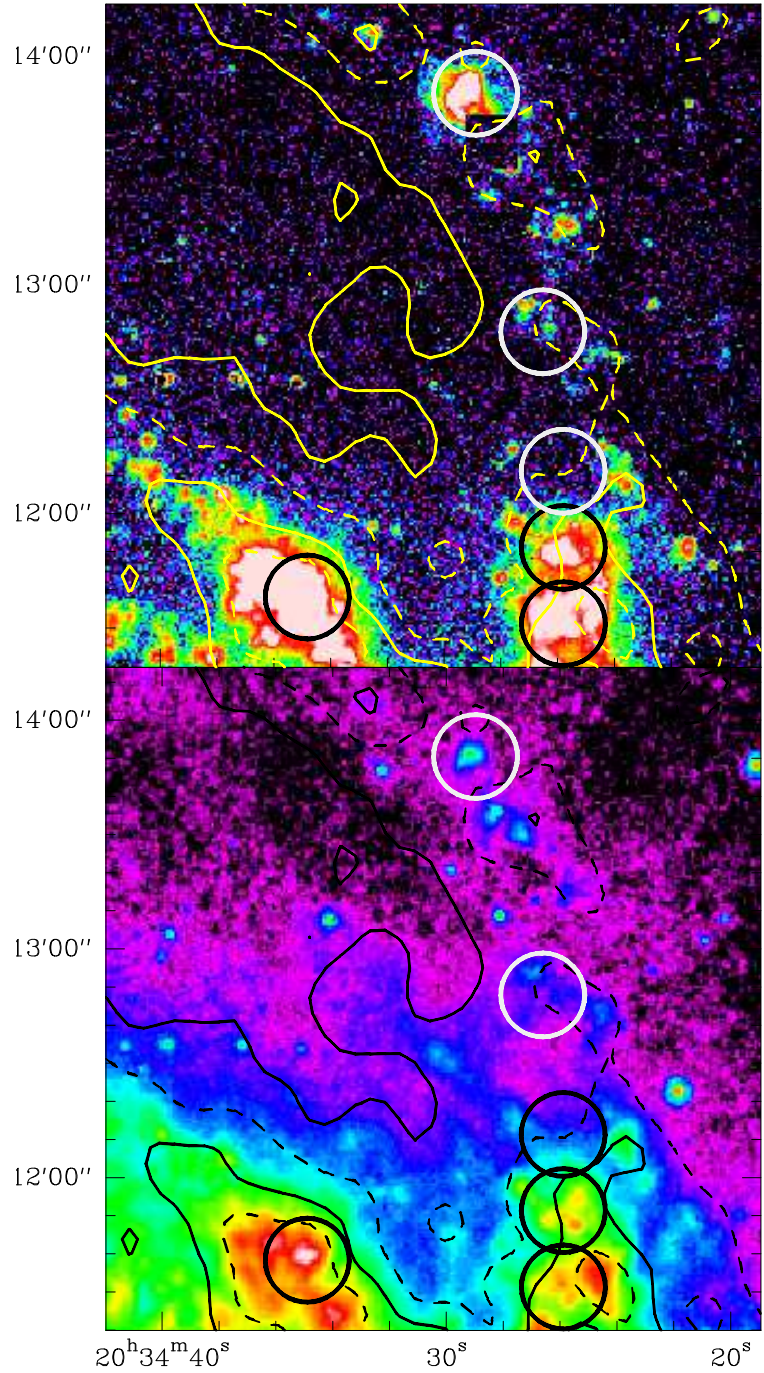}
\end{center}
\caption{(Top) An H$\alpha$ image of the spiral arm segment containing
   the P2, P7 and P9--12 positions, with the CO(1--0) beam superposed.
   The HI emission is shown as yellow contours at the 5, 10, 15, and 20
   $\times 10^{20}$ cm$^{-2}$ column density levels.  (Bottom) As above
   but overlaid on the Spitzer IRAC 8 $\mu$m image.
  \label{beams}}
\end{figure}

Figure~\ref{beams} shows a view of the spiral arm segment containing
positions P2, P7, and P9--P12 in the H$\alpha$ line and in 8$\mu$m
emission, usually attributed to PAHs, with HI contours overlaid. 
The PAH emission was observed as part
of the SINGS project with Spitzer.  While the locations of the HI, H$\alpha$, and PAH 8$\mu$m
emission are closely correlated, the intensities
of the emission at different wavelengths {\it do not} correlate. While
P11 and P2 have similar H$\alpha$ intensities (see also Table~1), the
8$\mu$m and CO emission is much weaker further out.  The
N(H$_2$)/N(HI) ratio also decreases, by
more than an order of magnitude, from 0.7 to 1.4 R$_{25}$ and is
largely driven by a sharp decrease in the CO intensity as one crosses
R$_{25}$ (P12 to P7).  

Table~1 also lists the implied H$_2$ column density at each position, derived
using a "standard" conversion factor of $\ratioo = 2 \times 10^{20}$
H$_2$ cm$^{-2}$ per K km s$^{-1}$ \citep{dickman86}, and the molecular
gas mass within the beam, including a correction for Helium. 
The masses inferred are in the range
$\sim2-70 \times 10^{5}$M~$_{\sun}$.  \cite{digel94} studied molecular
clouds in the far outer disk of the Milky Way (presumably beyond R$_{25}$ 
assuming a standard rotation curve) and found clouds of much
lower mass, in the range $2-40 \times 10^{3}$M~$_{\sun}$.  If such
clouds also exist in the outskirts of NGC~6946, they would
individually lie well below our detection limit of $\sim
10^{5}$M~$_{\sun}$.

\section{Discussion}

We have presented evidence for the existence of molecular gas in the
far outskirts of NGC~6946 and noted a rather abrupt decrease in
the molecular gas fraction, as well as the PAH emission, 
as one crosses R$_{25}$ along an individual
spiral arm.  An important question is whether this behavior could be
explained by metallicity variations in the underlying gas disk.
Indeed there is strong evidence for a radially increasing $\ratioo$
value \citep[e.g.][]{Sodroski95} as well as much higher values in
low-metallicity systems \citep[e.g.][]{Rubio91}.  The metallicity in the outer disk of
NGC~6946 varies by a factor $\sim 5$ from log(O/H)$=-3.2$ to $-3.9$ \citep{ferg98b}
while the I$_{\rm CO}$/N(HI) ratio decreases by a factor of $\sim30$.
Focusing only on the positions which bridge R$_{25}$ (P12 to P2), the
metallicity remains essentially constant while the I$_{\rm CO}$/N(HI) ratio
decreases by a factor of $\sim6$.  In the absence of a very strong
radiation field, the $\ratioo$ factor should not vary more than
linearly with metallicity and \citet{Wilson95} estimates that
$\ratioo \propto [O/H]^{2/3}$ from a study of molecular clouds in
Local Group galaxies.  It is thus reasonable to expect that the
$\ratioo$ ratio might be a factor 2 -- 3 higher in the most distant
points than at P9, where the "standard" value is probably
appropriate.  Such a correction is not sufficient to explain the
strong decrease in I$_{\rm CO}$/N(HI) at R$_{25}$ and we
conclude that there is a genuine drop in the molecular gas content at
the edge of the optical disk as defined by R$_{25}$.

Although our pointings were selected to lie on or near outer disk HII
regions, the H$\alpha$ luminosity contained within each beam shows
significant variation and corresponds to local star formation rates
(SFRs) ranging from $10^{-2}-10^{-4}$ M$_\odot$ yr$^{-1}$, using the
\citet{kenn98} calibration (SFR$=$L(H$\alpha / 1.26 \times 10^{41}$
erg s$^{-1}$). The H$\alpha$ emission in some regions is
predominantly diffuse (e.g. P12 and P7) and may more reflect
ionization by photons produced elsewhere rather than {\it in situ} star formation
\citep{ferg96}. The star formation efficiency (SFE), defined as SFR/M(H$_2$),
ranges from $\sim 10-0.07$ Gyr$^{-1}$; eliminating
the two outlier points (P12 and P2), the range reduces to $2 - 0.5$
Gyr$^{-1}$.  The mere presence
of a luminous HII region does not guarantee the detection of molecular
gas.  Position P1 in NGC~1058 is the second most luminous star-forming
region targeted in our study, and is our second most sensitive
integration, yet CO was not detected there.  Intriguingly, there is a
potential correlation between the detection of CO(2--1) emission and
the presence of a luminous HII region in that the three non-detections
of CO(2--1) are among the four lowest H$\alpha$ fluxes.

With the exception of P4, the HI column densities at our observed
positions within NGC~6946 are quite high, close to or greater than
$10^{21}$cm$^{-2}$ on the scale of several hundred parsecs.  The HI
column density may be an important factor in determining whether 
molecular gas
will be detected but a high HI column is clearly not a sufficient
condition \citep[see also][]{Gardan07}.  

It may be purely coincidental that N(H$_2$)/N(HI) changes abruptly
across R$_{25}$ in NGC~6946. While R$_{25}$ provides a means to
characterise the optical extents of galaxies, it is not expected to
relate to any underlying physical properties. In NGC~6946, neither the
metallicity, the HI surface density nor the surface brightness of the
stellar disk exhibit unusual features at this
location.  Interestingly, the azimuthally-averaged areal star formation
rate in NGC~6946 does show a sharp decline near the edge of
the optical disk ($\sim 0.8$~R$_{25}$), although widespread low-level
star formation is observed much further out \citep{ferg98a,mk01}.  The
rough correspondence between the star formation and molecular profile
breaks could provide support for threshold models
\citep[e.g.][]{elme94,schaye04} although the physics that drives this
behavior is still open to debate.  Pressure based \citep{elme94,blitz06}
and column density based \citep{schaye04} models are very difficult to
distinguish once the HI surface density exceeds that of the stars.
Larger-scale mapping of the outer
disk of NGC~6946 and sensitive observations of the outskirts of
additional galaxies are required in order to more thoroughly
understand the molecular content and distribution at large
galactocentric radii.

\acknowledgments AMNF acknowledges a Marie Curie Excellence Grant from
the European Commission under contract MCEXT-CT-2005-025869 and a
Visiting Professorship at the Observatoire de Bordeaux where
this work was completed.  We thank Rense Boomsma and Tom Oosterloo for
kindly sharing their HI data on NGC~6946 prior to publication.

\begin{deluxetable}{lccccccccccccc}
\tablecolumns{14}
\tabletypesize{\scriptsize}
\rotate
\tablecaption{Molecular Gas in the Extended Disk of NGC~6946\label{table1}}
\tablewidth{0pt}
\tablehead{
\colhead{Posn.} & \colhead{R.A.} & \colhead{Dec.} & \colhead{R/R$_{25}$}  & \colhead{I$_{\mathrm CO(1-0)}$} & \colhead{rms} & \colhead{$\Delta$V} & \colhead{I$_{\mathrm CO(2-1)}$} & \colhead{rms} & \colhead{(O/H)} & \colhead{N(H$_2$)} & \colhead{M(H$_2$)} & \colhead{N(HI)} & \colhead{L(H$\alpha$)}\\
   & \colhead{(J2000.0)} & \colhead{(J2000.0)}  & & \colhead{K~km~s$^{-1}$} & \colhead{mK} &  \colhead{km~s$^{-1}$} & \colhead{K~km~s$^{-1}$} & \colhead{mK} & \colhead{dex} & \colhead{10$^{20}$cm$^{-2}$} & \colhead{10$^5$M$\sun$} & \colhead{10$^{20}$cm$^{-2}$} & \colhead{10$^{37}$erg~s$^{-1}$}\\
\colhead{(1)} & \colhead{(2)} &\colhead{(3)} &\colhead{(4)} &\colhead{(5)} &\colhead{(6)} &\colhead{(7)} &\colhead{(8)} &\colhead{(9)} &\colhead{(10)} &\colhead{(11)} &\colhead{(12)} &\colhead{(13)} &\colhead{(14)} }
\startdata
\cutinhead{NGC~6946}
P9  & 20 34 34.9 & 60 11 38.4 & 0.7  & 4.5$\pm .1$     & 17  & 19    & 6.3$\pm .3$   & 39    & -3.2 & 9.0    &70 & 22.1 & 142.5\\
P10 & 20 34 25.9 & 60 11 31.3 & 0.8  & 2.3$\pm .1$     & 16  & 22    & 2.2$\pm .2$   & 30    &  -3.3 & 4.6    & 36 & 17.1 & 64.3 \\
P11 & 20 34 25.9 & 60 11 51.3 & 0.8  & 2.2$\pm .1$     & 15  & 24    &  3.8$\pm .3$  & 32    &  -3.3  & 4.5    & 34 & 15.2 & 28.9 \\
P12 & 20 34 25.9 & 60 12 11.3 & 0.9  & 1.0$\pm .1$    & 10  & 30    & $<0.5$& 23    &  -3.4 & 2.0    & 15 & 10.1 & 1.4 \\
P7  & 20 34 26.6 & 60 12 47.9 & 1.0  & 0.21$\pm .03$    & 4.3   & 18    & 0.22$\pm .05$ & 7     &   -3.4 & 0.4    & 3.3 & 9.2  & 2.2  \\
P2  & 20 34 29.0 & 60 13 50.5 & 1.1  & 0.14$\pm .02$    & 3.0   & 14    & 0.14$\pm .03$ & 5     &  -3.5  & 0.3    & 2.2 & 8.3  & 28.8 \\
P8  & 20 34 54.3 & 60 15 40.5 & 1.3  & 0.15$\pm .02$    & 5.0   & 8     & $<0.3$ & 13    &  -3.6 & 0.3    & 2.3   & 8.9  & 4.2  \\
P3  & 20 33 48.4 & 60 08 32.6 & 1.4  & 0.11$\pm .02$ & 3.7   & 12 & $<0.13$& 8
   &  -3.9 & 0.2 & 1.7   & 13.6 & 4.7  \\
P5  & 20 34 51.4 & 60 16 24.9 & 1.5  & $<.13$  & 6.1  & \nodata & $<0.5$ & 24    &  -3.8 & $<.3$&$<1.9$   & 10.2 & $<0.7$  \\
P4  & 20 34 51.4 & 60 16 44.9 & 1.6  & $<0.14$ & 6.7   & \nodata & $<0.4$ & 20    & -3.8  & $<.3$& $<2.0$    & 5.0  & $<0.1$ \\
\cutinhead{NGC~1058}
P1  & 02 43 29.3 & 37 18 57.6 & 1.0  & $<0.07$ & 3.5   & \nodata & \nodata & \nodata & -3.3    & $<0.14$&\nodata    & 3.7 & 85.5 \\
P2  & 02 43 29.4 & 37 17 40.0 & 1.9  & $<0.11$ & 5.4   & \nodata & \nodata & \nodata & -3.8    & $<0.21$& \nodata   & 4.2 & 11.7\\
\enddata
\tablecomments{Units of right ascension are hours, minutes and seconds
   and units of declination are degrees, arcminutes and
   arcseconds. Col.(1): position ID. Cols.(2) and (3): Right ascension
   and declination of the pointing. Col.(4) the deprojected radius of
   the pointing in terms of R$_{25}$, calculated for NGC~6946 using
   R$_{25}=5.75\arcmin$ \citep{rc3}, and $i = 34.0^{\circ}$ and
   P.A.$=69^{\circ}$ \citep{ca88} and for the almost face-on NGC~1058
   using R$_{25}=1.51\arcmin$ \citep{rc3} and $i = 0^{\circ}$ and
   P.A.$=0^{\circ}$. Cols.(5) and (8): the CO(1--0) and CO(2--1)
   intensities on the main beam temperature scale respectively. Upper
   limits are 3$\sigma$ and calculated assuming a 7 channel (18.2
   km~s$^{-1}$) line width. Cols.(6) and (9): the rms noise, 
   measured per 2.6 km~s$^{-1}$ channel. Col.(7): the
   velocity width as determined from fitting gaussians to the CO(1--0)
   spectra.  Col.(10) The oxygen abundance, from (\citet{ferg98b} and
   Ferguson, unpublished. Note that the metallicities for
   NGC~6946-P3 and NGC~1058-P1 and P2 are directly measured, whereas
   those for other positions are inferred from interpolation of the
   abundance gradient. Cols. (11) and (12) the H$_2$ column density and
   cloud mass, as described in the text. Col.(13) The HI column density at the position
   from \citet{boomsma07} and \citet{dickey90}. Col.(14) The H$\alpha$
   luminosity within a 21$\arcsec$ aperture, corrected for Galactic
   extinction and [NII] emission.}
\end{deluxetable}

\clearpage

\begin{thebibliography}{}
\bibitem[Blitz \& Rosolowsky(2006)]{blitz06} Blitz, L., \& Rosolowsky, E.\ 2006, \apj, 650, 933
\bibitem[Boomsma(2007)]{boomsma07} Boomsma, R.\ 2007, Ph.D.~Thesis, University of Groningen.
\bibitem[Braine \& Herpin(2004)]{braine04} Braine, J., \& Herpin, F.\ 2004, \nat, 432, 369
\bibitem[Braine et al.(2001)]{Braine01} Braine, J., Duc, P.-A., \& Lisenfeld, U. et al.\ 2001, \aap, 378, 51
\bibitem[Brouillet et al.(1992)]{brou92} Brouillet, N., Henkel, C., \& Baudry, A.\ 1992, \aap, 262, L5
\bibitem[Considere \& Athanassoula(1988)]{ca88} Considere, S., \& Athanassoula, E.\ 1988, \aaps, 76, 365
\bibitem[de Vaucouleurs et al.(1991)]{rc3} de Vaucouleurs, G., de Vaucouleurs, A., Corwin, H.~G., Jr., Buta, R.~J., Paturel, G., \& Fouque, P.\ 1991, Volume 1-3, XII, 2069 pp.~7
\bibitem[Dickey et al.(1990)]{dickey90} Dickey, J.~M., Hanson, M.~M., \& Helou, G.\ 1990, \apj, 352, 522
\bibitem[Dickman et al.(1986)]{dickman86} Dickman, R.~L., Snell, R.~L., \& Schloerb, F.~P.\ 1986, \apj, 309, 326
\bibitem[Digel et al.(1994)]{digel94} Digel, S., de Geus, E., \& Thaddeus, P.\ 1994, \apj, 422, 92
\bibitem[Dobbs et al.(2006)]{dobbs06} Dobbs, C.~L., Bonnell, I.~A., \& Pringle, J.~E.\ 2006, \mnras, 371, 1663
\bibitem[Elmegreen \& Parravano(1994)]{elme94} Elmegreen, B.~G., \& Parravano, A.\ 1994, \apjl, 435, L121
\bibitem[Ferguson et al.(1998a)]{ferg98a} Ferguson, A.~M.~N., Wyse, R.~F.~G., Gallagher, J.~S., \& Hunter, D.~A.\ 1998a, \apjl, 506, L19
\bibitem[Ferguson et al.(1998b)]{ferg98b} Ferguson, A.~M.~N., Gallagher, J.~S., \& Wyse, R.~F.~G.\ 1998b, \aj, 116, 673
\bibitem[Ferguson et al.(1996)]{ferg96} Ferguson, A.~M.~N., Wyse, R.~F.~G., Gallagher, J.~S., III, \& Hunter, D.~A.\ 1996, \aj, 111, 2265
\bibitem[Gardan et al.(2007)]{Gardan07} Gardan, E., Braine, J., Schuster, K.~F., Brouillet, N. \& Sievers, A.\ 2007, \aap, 473, 91
\bibitem[Kennicutt(1998)]{kenn98} Kennicutt, R.~C., Jr.\ 1998, \apj, 498, 541
\bibitem[Kim et al.(2002)]{kim02} Kim, W.-T., Ostriker, E.~C., \& Stone, J.~M.\ 2002, \apj, 581, 1080
\bibitem[Martin \& Kennicutt(2001)]{mk01} Martin, C.~L., \& Kennicutt, R.~C., Jr.\ 2001, \apj, 555, 301
\bibitem[Regan et al.(2006)]{regan06} Regan, M.~W., et al.\ 2006, \apj, 652, 1112
\bibitem[Pfenniger et al.(1994)]{Pfeniger94} Pfenniger, D., Combes, F., Martinet, L.\ 1994, \aap, 285, 79
\bibitem[Rubio et al.(1991)]{Rubio91} Rubio, M., Garay, G., Montani, J., \& Thaddeus, P.\ 1991, \apj, 368, 173
\bibitem[Ryan-Weber et al.(2004)]{ryanw04} Ryan-Weber, E.~V., et al.\ 2004, \aj, 127, 1431
\bibitem[Schaye(2004)]{schaye04} Schaye, J.\ 2004, \apj, 609, 667
\bibitem[Sodroski et al.(1995)]{Sodroski95} Sodroski, T.~J., Odegard, N., Dwek, E., et al.\ 1995, \apj, 452, 262
\bibitem[Skillman (1987)]{skill87} Skillman, E.~D.\ 1987, in Star Formation in Galaxies, p. 263.
\bibitem[Thilker et al.(2005)]{thilker05} Thilker, D.~A., et al.\ 2005, \apjl, 619, L79
\bibitem[Walter et al.(2006)]{walter06} Walter, F., Martin, C.~L., \& Ott, J.\ 2006, \aj, 132, 2289
\bibitem[Wilson(1995)]{Wilson95} Wilson, C.~D. 1995, \apjl, 448, L97
\end{thebibliography}
\end{document}